# Compact optical phased array using a serial grating antenna design

Lingxuan Zhang[1, 2], Xiaochen Sun[1, 2,#], Wenfu Zhang[1, 2,*], Guoxi Wang[1, 2], Ningning Feng[1, 2], and Wei Zhao[1, 2]

[1]State Key Laboratory of Transient Optics and Photonics, Xi'an Institute of Optics and Precision Mechanics, Chinese Academy of Sciences, Xi'an 710119, China
[2]University of Chinese Academy of Sciences, No.19A Yuquan Road, Beijing 100049, China
E-mail: *wfuzhang@opt.ac.cn, #sunxiaochen@opt.ac.cn

## Abstract

We propose an on-chip Si photonics optical phased array based on a serial grating design which eliminates the use of directional couplers in previous designs. It significantly reduces overall phased array size especially when the number of the antenna is small which is often demanded in practice. The simulation results show our design reduces overall phased array size, increases optical power utilization while maintains comparable far field performance.

## Introduction

Radio wave phased arrays play important roles in modern communication, ranging and astronomy [1]. Based on the same physics but a drastically different frequency range, chip scale silicon photonics optical phased array has lately been drawing increasing attention for a wide range of applications from free-space communication [2] to image projection [3]. A phased array device is generally made of many antenna units which are arranged in 2D array and are individually tunable in phase in order to form a specific output beam pattern through interference effect. The spacing between antenna units in a radio wave phased array is usually well below the operating radio wavelength to reduce higher order interference, however, in an optical phased array it is generally much larger than the optical wavelength due to the limitation of optical waveguide design and fabrication. Then it is essential to make these antenna units as close to each other as possible to reduce interference hence side lobes in output beam [4-6].

The schematic drawing of a recent phased array design is shown in Fig. 1 [7, 8]. And each antenna unit is composed of three elements: an antenna, a phase shifter and a directional coupler. In such design, the minimal spacing between adjacent antenna units is generally limited by the length of the directional couplers as the adjustment in the sizes of the grating antenna and the phase shifter are very limited for a given wavelength and a manufacturing process.

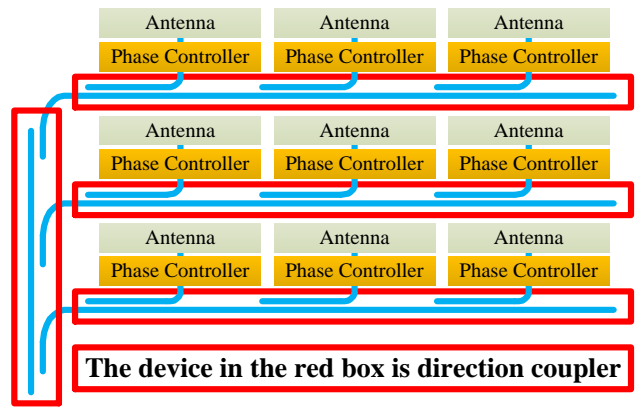

Fig. 1 Schematic drawing of a recent optical phased array design.

For example, considering a phased array with 9 by 9 antenna units and a reasonable 81% optical power utilization made on a common 220nm SOI platform with a critical dimension of 180nm, it requires at least a total length of 71µm for 9 directional couplers in a row while the combined size of a grating antenna and a phase shifter (realized with a S-bend as in [7, 8]) in the row can be as short as 48.6µm. The power utilization here is defined as the theoretical percentage of optical power that is coupled to all grating antennas while waveguide propagation loss and upward diffraction loss are not counted for fair comparison between different designs. A more demanding optical power utilization (e.g. 98%, on par with the number our design achieves) would require even longer directional couplers (e.g. 95µm). Utilizing large number of antennas could allow smaller directional couplers for less percentage of power tapping at each one, however, it would render proportionally more complex electronic controlling circuits, more I/O counts, and most importantly more power consumption which may prevent its use in practice. Reducing optical power utilization could also reduce the length of directional couplers however again at the cost of power consumption given the same total output power.

In this paper, we introduce a serial shallow-etched grating design to totally eliminate the use of directional couplers therefore to solve this dilemma. The shallow-etched gratings are connected in series by optical waveguides while arranged in 2D array and they serve both as antennas and as optical power taps. As a result, the spacing between antennas can be small and is independent of the number of antenna units as well as optical power utilization. The reduction in antenna spacing not only reduces device footprint but also helps minimize side lobes in output beam. Although it introduces unequal output optical power at different antennas, we confirm it shows insignificant impact on performance.

## Design

A block diagram and a 3D schematic representation of the presented design are shown in Fig. 2. A straight grating antenna, instead of a focusing one as used in [7, 8], is used and it allows a small part of the light scattered out of plane at a design angle while allows the rest part of light to pass through. A phase shifter, designed as an S-bend, is placed between two adjacent grating antenna connected with waveguides. The two branches of a phase shifter circled in green in Fig. 2 (b) are where electrodes contact the lightly doped Si waveguide in between to change its phase by heating it up. The grating antenna (i.e. circled in blue) and the phase shifter are closely packed in a multi-section serpentine shape to reduce the overall size. The distance in x-axis between two 1D arrays of antenna units can be designed by adjusting the length of the waveguide section circled in red in order to introduce aperiodic array design to reduce side lobes [9].

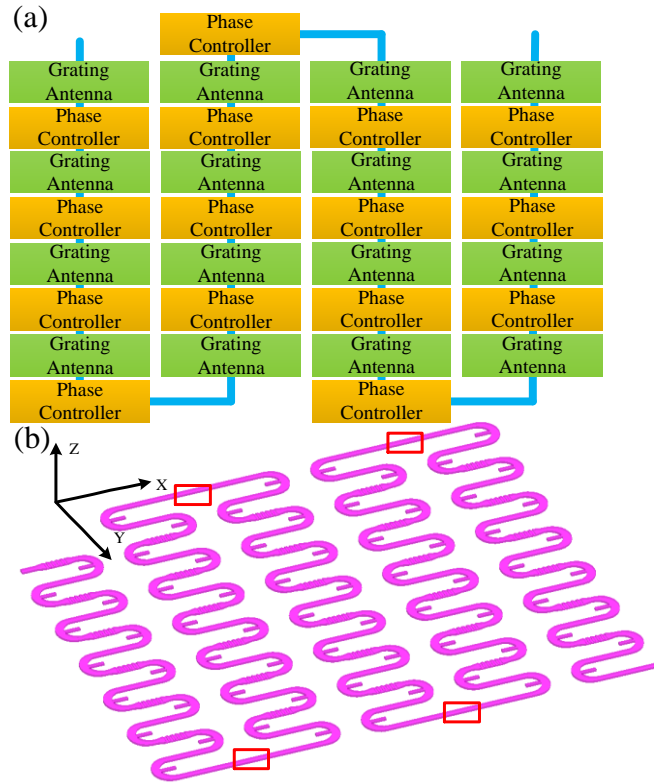

Fig. 2 (a) A block diagram and (b) a 3D schematic representation of the presented design. The grating antenna is circled in blue, while the two branches for electrode contact are circled in green.

The 3D structure of the grating antenna is shown in Fig. 3 (a). The grating is made by Si as core and $SiO_2$ as cladding and a thin layer of silver is added at the bottom of the grating to prevent downward energy leakage therefore to enhance the upward diffraction and pass-through efficiencies. The efficiencies of pass-through transmission $T$ and upward diffraction $D$ with respect to tooth height $h$ are calculated by finite-difference time-domain (FDTD) approach [11-13] and are shown in Fig. 3 (c). By choosing the period of grating antenna at 0.62μm, corresponding to a 13.6° output angle at 1.55μm wavelength, and the tooth height $h$ at 20 nm, we can obtain pass-through transmission of 95% and upward diffraction of 5% which are used for the rest of the work. A simulated Gaussian-like far field pattern of a grating with 5 periods and a width of 1 um are shown in Fig. 3 (b). The diffracted beam is centered at 13.6° with -4.5° and 25.3° half divergence angles (defined at $1/e^2$ of peak intensity) in x and y, respectively. For a 9×9 antenna array design a 98.4% total optical power utilizing is achieved with an approximate length of 66μm in both x and y directions which is a 30% linear dimension reduction compared to the previous design described earlier (i.e. 95μm). The design can simply be optimized by adjusting grating tooth height for other array sizes and optical power utilization.

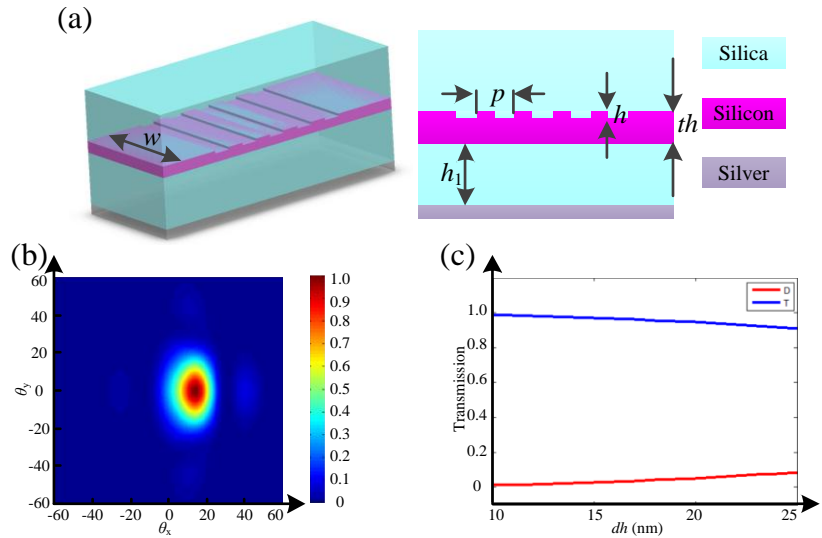

Fig. 3 (a) 3D structure of the grating antenna with following parameters: $w$=1μm, $p$=0.62μm, $th$=220nm, $h$=20nm and $h_1$=1μm. (b) Normalized Gaussian-like far-filed pattern of the grating antenna. (c) Pass-through transmission $T$ and upward diffraction $D$ v.s. tooth height.

An elliptical S-bend phase shifter described in [7, 8, 10] is adopted in the design. A typical design such S-bend with design parameters shown in Fig. 4 (a) can theoretically achieve transmission as high as 99% as shown in Fig. 4 (b). As mentioned earlier, it demands adjusting the S-bend parameter $R_{b1}$ in order to purposefully introduce aperiodicity. The loss-optimized values of other parameters (i.e. $R_{b2}$, and $R_{a2}$) at different $R_{b1}$ are shown in Fig. 4 (c).

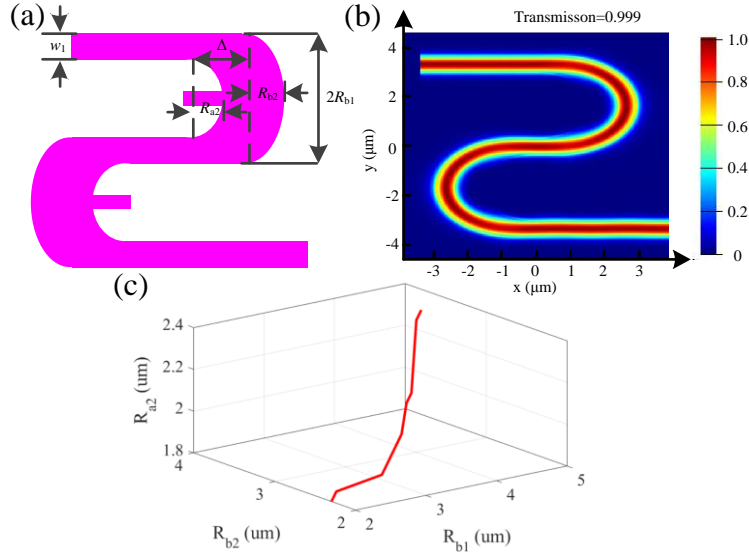

Fig. 4 (a) Design and (b) FDTD-simulated optical propagation of an elliptical S-bend with following parameters: $R_{b1}$=2μm, $R_{b2}$=2.3μm, $R_{a2}$=1.8μm, Δ=0.6μm, $w_1$=0.65μm. (c) Loss-optimized $R_{b2}$, and $R_{a2}$ v.s. $R_{b1}$ with Δ=0.6μm

## Effect of unequal optical power of antennas

As all the grating antennas in this design are likely the same considering using a single Si etch step in practice, it brings up an issue of unequal output optical power of each antenna. Changing the number of grating period could change diffraction property however such change is limited

(e.g. from a baseline of 5 periods) and too discrete to make fine tuning. Therefore it is important to study the effect of such antenna output power inequality on the performance of phased array. We use a phased array with 9×9 8μm-spaced, evenly distributed antenna units operating at 1.55 um wavelength as an example. The output of each antenna is expressed as

$$E_{pq} = \begin{cases} T^{9(q-1)+(p-1)}D & \mod(q,2)=1 \\ T^{9(q-1)+(9-p)}D & \mod(q,2)=0 \end{cases}, \quad (1)$$

, where $T$ indicates pass-through transmission and $D$ indicates diffracted optical power of a single grating antenna. Then the far field can be calculated as

$$E = \sum_{p,q} E_{pq} e^{i\varphi_{pq}}, \quad (2)$$

The phase $\varphi_{pq}$ is expressed as

$$\varphi_{pq} = \frac{(p-1)p_x \tan(\theta_x) + (q-1)p_y \tan(\theta_y)}{\sqrt{1+\tan^2(\theta_x)+\tan^2(\theta_y)}}. \quad (3)$$

, where $p_{x(y)}$ is the period in x (y) axis, and $\vartheta_{x(y)}$ is diffractive angle in x(y) axis, respectively. We calculate far field $E$ with $T$=0.95 and $D$=0.05 and the results are shown in Fig. 5 (a)-(c). As a comparison, the results of an ideal phased array with equal optical power of each antenna are shown in Fig. 5 (d)-(f). Although it can be found that the phased array with unequal optical power brings 6% noise by comparing Fig. 5 (b) and (e), impact of the noise is nearly the same of the side lobe shown in Fig. 5 (e). Therefore, this unequal antenna output effect is acceptable for ordinary application of phased array.

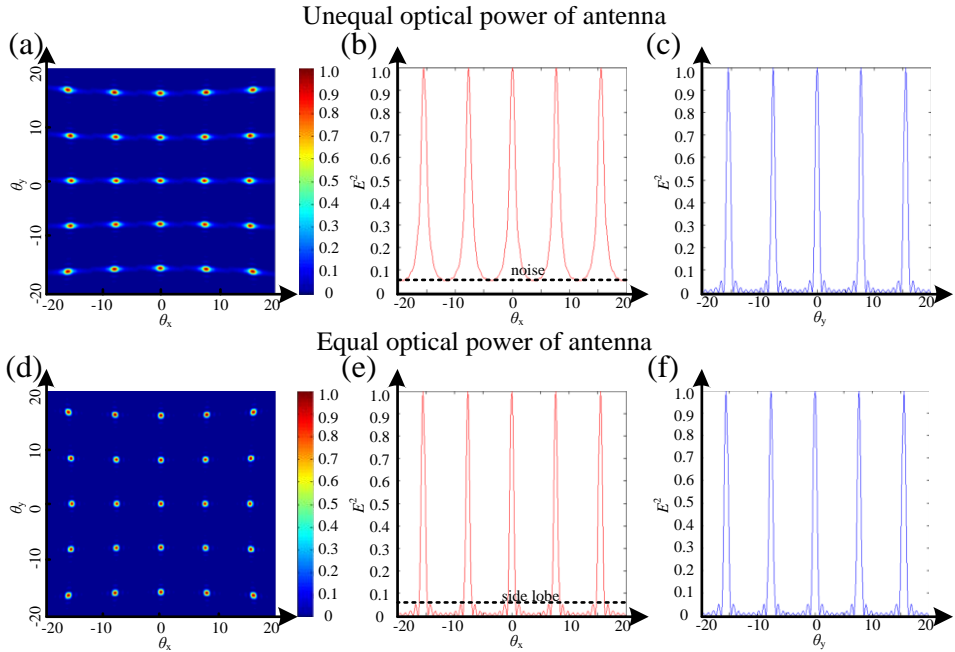

Fig. 5 (a)-(c) Far field results of the phased array with unequal antenna output. (d)-(f) Far field results of the phased array with equal antenna output.

## Discussion on size

It has been pointed out in the introduction that the size of phased array shown in Fig. 1 is dominated by the length of directional couplers when the number of antenna is small. In comparison, the proposed design in this paper eliminates directional couplers and can significantly reduce overall phased array size in many cases. In this section, we further analyze the effect of antenna number on size reduction. In the following analysis, a total optical power unitization $R^2$ is always assumed (e.g. 90%). The power utilization of each branch is then R and for an $N\times N$ antenna phased array in Fig. (1), we can obtain

$$R = N\eta_1, \tag{4}$$

, where $\eta_1$ is coupling efficiency of the first directional coupler, and coupling efficiency of other directional coupler can be obtained as

$$\eta_n = \frac{\eta_1}{\prod_{i=1}^{n-1}(1-\eta_i)}, \tag{5}$$

Accordingly, the total length, $L_{tot}$ of an antenna branch can be obtained as

$$L_{tot} = \sum_{i=1}^{N} L_i, \tag{6}$$

, where $L_i$ is size of each antenna branch, and $L_i$ can be calculated as

$$L_i = \begin{cases} L_c(\eta_i) & L_c(\eta_i) > L_a \\ L_a & L_c(\eta_i) < L_a \end{cases}, \tag{7}$$

, with $L_a$ is the size of the antenna (including the S-bend phase shifter), then the average size of an antenna unit, $L_{ave}$, is

$$L_{ave} = \frac{L_{tot}}{N}, \tag{8}$$

The coupling length $L_c$ with respect to coupling efficiency $\eta$ of a directional coupler is calculated based on a commonly used waveguide cross section of 500nm×220nm [13-15] and a gap of 180 nm [14, 16] and the result is shown in Fig. 6 (a). $L_{ave}(N)$ for different $R^2$ with respect to $N$ is calculated and shown in Fig. 6 (b). It demonstrates that the average size of antenna unit, $L_{ave}$, in the previous design increases significantly with the number of antenna $N$ get smaller (and more practically favorable). At a given $R^2$ (0.98 in this example), the $L_{ave}$ in our design is independent of $N$, and can significantly reduce size in such cases.

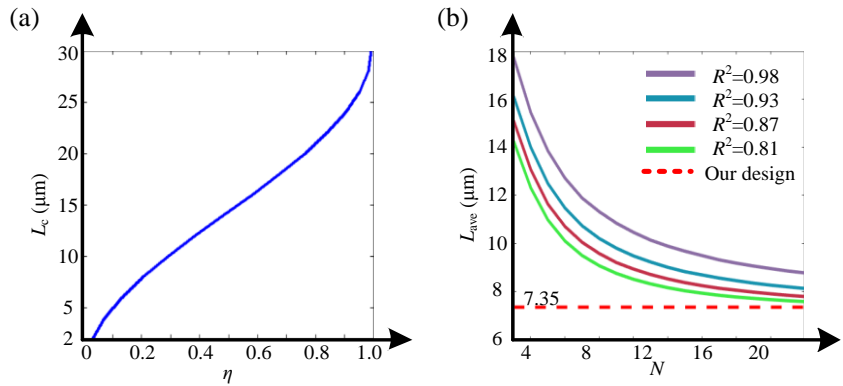

Fig. 6 (a) Coupling length $L_c$ v.s. coupling efficiency $\eta$ of a directional coupler. (b) Average size of antenna unit $L_{ave}$ at different total optical power utilization $R^2$ v.s. number of antenna $N$ in a branch. The average size of an antenna unit of our design is $L_a$=7.35µm with $R^2$ of 0.98.

## Conclusion

In this paper, we introduce a Si photonics optical phased array based on a serial grating design that eliminates the use of directional couplers in previous designs and can significantly reduce overall phased array size especially when the number of the antenna is small which is often demanded in practice. The simulation results show our design reduces overall phased array size, increases optical power utilization while maintains comparable far field performance

## Acknowledgements

This work was supported by National Natural Science Foundation of China (Grant No. 11404388, 61405243, 61475188 and 61705257), the Strategic Priority Research Program of the Chinese Academy of Sciences (Grant No. XDB24030600).